\documentclass[onecolumn,epjc3]{svjour3}

\RequirePackage{fix-cm}
\smartqed
\RequirePackage{graphicx}
\RequirePackage{subfig}
\RequirePackage{float}
\usepackage{rotating}
\usepackage{amsmath,amssymb}
\usepackage{mathtools}
\RequirePackage[colorlinks,citecolor=blue,urlcolor=blue,linkcolor=blue]{hyperref}
\usepackage{doi}
\usepackage[]{units}
\usepackage{xspace}
\usepackage[switch]{lineno}
\usepackage{cite}
\usepackage{comment}
\usepackage{url}
\usepackage[version=4]{mhchem}
\usepackage{upgreek}
\usepackage{booktabs}
\usepackage{bold-extra}
\usepackage{lineno}

\newcommand{\cuore}{CUORE}
\newcommand{\cuoreo}{CUORE-0}

\newcommand{\cupid}{CUPID}
\newcommand{\cupido}{CUPID-0}
\newcommand{\cupidmo}{CUPID-Mo}

\newcommand{\dbd}{$2\mathrm{\nu\beta\beta}$}

\newcommand{\ndbd}{$0\mathrm{\nu\beta\beta}$}
\newcommand{\nnbd}{$2\mathrm{\nu\beta\beta}$}

\newcommand{\mbb}{$m_{\beta\beta}$}
\newcommand{\Qbb}{$Q_{\beta\beta}$}
\newcommand{\thalf}{$T_{1/2}^{0\nu}$}

\newcommand{\Mo}{\ce{^{100}Mo}}
\newcommand{\Te}{\ce{^{130}Te}}
\newcommand{\Se}{$^{82}$Se}
\newcommand{\Cd}{$^{116}$Cd}
\newcommand{\Ca}{$^{48}$Ca}
\newcommand{\Ge}{$^{76}$Ge}
\newcommand{\Xe}{$^{136}$Xe}
\newcommand{\LMO}{\ce{Li$_2$MoO$_4$}}
\newcommand{\lmo}{\LMO}
\newcommand{\enrlmo}{\ce{Li_{2}^{{100}}MoO_{4}}}

\newcommand{\enrcwo}{\ce{$^{116}$CdWO_{4}}}

\newcommand{\enrZS}{\ce{Zn$^{82}$Se}} 

\newcommand{\Tl}{$^{208}$Tl}

\newcommand{\Th}{$^{232}$Th}
\newcommand{\U}{$^{238}$U}

\newcommand{\Bi}{$^{214}$Bi}

\newcommand{\bega}{$\beta/\gamma$}
\newcommand{\A}{$\alpha$}
\newcommand{\B}{$\beta$}
\newcommand{\G}{$\gamma$}
\newcommand{\BG}{$\beta/\gamma$}

\newcommand{\ckky}{cts$/($keV$\cdot$kg$\cdot$yr$)$}

\journalname{Eur. Phys. J. C} 
\begin{document}

\title{CUPID, the \textsc{Cuore} Upgrade with Particle IDentification}

\author{K.~Alfonso\thanksref{VT_US,UCLA_US}
\and
A.~Armatol\thanksref{LBNL_US,fn0}
\and
C.~Augier\thanksref{IP2I_France}
\and
F.~T.~Avignone~III\thanksref{UofSC_US}
\and
O.~Azzolini\thanksref{LNL_Italy}
\and
A.S.~Barabash\thanksref{NRC_KI_Russia}
\and
G.~Bari\thanksref{SdB_Italy}
\and
A.~Barresi\thanksref{UniMIB_Italy,MIB_Italy}
\and
D.~Baudin\thanksref{CEA_IRFU_France}
\and
F.~Bellini\thanksref{SdR_Italy,SURome_Italy}
% ...
}

\institute{Virginia Polytechnic Institute and State University, Blacksburg, VA, USA\label{VT_US}
\and
University of California, Los Angeles, CA, USA\label{UCLA_US}
\and
Lawrence Berkeley National Laboratory, Berkeley, CA, USA\label{LBNL_US}
\and
Univ Lyon, Universit\'e Lyon 1, CNRS/IN2P3, IP2I-Lyon, Villeurbanne, France\label{IP2I_France}
\and
University of South Carolina, Columbia, SC, USA\label{UofSC_US}
\and
INFN Laboratori Nazionali di Legnaro, Legnaro, Italy\label{LNL_Italy}
\and
National Research Centre Kurchatov Institute, Kurchatov Complex of Theoretical and Experimental Physics, Moscow, Russia\label{NRC_KI_Russia}
\and
INFN Sezione di Bologna, Bologna, Italy\label{SdB_Italy}
\and
Universit\`a degli Studi di Milano-Bicocca, Dipartimento di Fisica, Milano, Italy\label{UniMIB_Italy}
\and
INFN Sezione di Milano-Bicocca, Milano, Italy\label{MIB_Italy}
\and
IRFU, CEA, Universit\'e Paris-Saclay, Saclay, France\label{CEA_IRFU_France}
\and
INFN Sezione di Roma, Rome, Italy\label{SdR_Italy}
\and
Sapienza University of Rome, Rome, Italy\label{SURome_Italy}
% ...
\and
INFN Laboratori Nazionali del Gran Sasso, Assergi (AQ), Italy\label{LNGS_Italy}
\and
INFN Laboratori Nazionali di Frascati, Frascati, Italy\label{LNF_Italy}
\and
Universit\'e Paris-Saclay, CNRS/IN2P3, IJCLab, Orsay, France\label{IJCLab_France}
\and
CNR-Institute for Microelectronics and Microsystems, Bologna, Italy\label{CNR-IMM_Italy}
\and
INFN Sezione di Pavia, Pavia, Italy\label{PV_Italy}
% ...
\and
University of California, Berkeley, Berkeley, CA, USA\label{UCB_US}
\and
INFN Sezione di Genova, Genova, Italy\label{SdG_Italy}
% ...
\and
Rome Tor Vergata University, Rome, Italy\label{Rome_Tor_Vergata_University_Italy}
\and
INFN sezione di Roma Tor Vergata, Rome, Italy\label{INFN_Tor_Vergata_Italy}
\and
Northwestern University, Evanston, IL, USA\label{NWU_US}
\and
Argonne National Laboratory, Argonne, IL, USA\label{ANL_US}
% ...
\and
Massachusetts Institute of Technology, Cambridge, MA, USA\label{MIT_US}
\and
Boston University, Boston, MA, USA\label{BU_US}
\and
California Polytechnic State University, San Luis Obispo, CA, USA\label{CalPoly_US}
% ...
\and
Yale University, New Haven, CT, USA\label{Yale_US}
\and
Department of Physics and Astronomy, University of Pittsburgh, Pittsburgh, PA, USA\label{Pittsburgh_US}
\and
Drexel University, Philadelphia, PA, USA\label{Drexel_US}
\and
Johns Hopkins University, Baltimore, MD, USA\label{JHU_US}
% ...
\and
Nikolaev Institute of Inorganic Chemistry, Novosibirsk, Russia\label{NIIC_Russia}
\and
INFN Sezione di Padova, Padova, Italy\label{PD_Italy}
\and
Univ. Grenoble Alpes, CNRS, Grenoble INP, SIMAP, Grenoble, France\label{SIMaP_Grenoble_France}
\and
University of Bologna, Bologna, Italy\label{UnivBologna_Italy}
}

\thankstext{fn0}{Now at IP2I-Lyon, Univ Lyon, France}
\thankstext{fn1}{Now at Physik-Institut, University of Z\"urich, Z\"urich, Switzerland}
\thankstext{fn2}{Now at University College London, London, UK}
\thankstext{fn3}{Also at Instituto de F{\'\i}sica, Universidade de S\~ao Paulo, Brazil and Max-Planck-Institut f\"ur Physik, M\"unchen, Germany}


\date{Received: date / Accepted: date}
\maketitle

\begin{abstract}
  \cupid, the CUORE Upgrade with Particle IDentification, is a next-generation experiment to search for neutrinoless double beta decay (\ndbd) and other rare events using enriched  \enrlmo\ scintillating bolometers. It will be hosted by the CUORE cryostat located at the Laboratori Nazionali del Gran Sasso in Italy.
  The main physics goal of \cupid\ is to search for \ndbd\ of \Mo\ with a discovery sensitivity covering the full neutrino mass regime in the inverted ordering scenario, as well as the portion of the normal ordering  regime with lightest neutrino mass larger than 10~meV. With a conservative background index of 10$^{-4}$~\ckky, 240~kg isotope mass, 5~keV FWHM energy resolution at 3 MeV and 10 live-years of data taking, \cupid\ will have a 90\% C.L. half-life exclusion sensitivity of $1.8\cdot10^{27}$~yr, corresponding to an effective Majorana neutrino mass (\mbb)  sensitivity of 9--15~meV,
  and a $3\sigma$ discovery sensitivity of $1\cdot10^{27}$~yr,
  corresponding to an \mbb\ range of 12--21~meV. 

\end{abstract}
\sloppy

\section{Physics case}

Double beta decay with two anti-neutrino emission (\dbd), $(A,Z) \to (A,Z+2)+2e^-+2\bar\nu$, is among the rarest nuclear transitions ever observed. It is a second-order weak transition allowed for 35 candidate even-even nuclei, where the decay to the intermediate nucleus is forbidden due to energy conservation or suppressed by a large change of the nuclear spin. This Standard Model process conserves lepton number and has been observed in eleven nuclei with half-lives in the range of 10$^{18}$--10$^{24}$~yr~\cite{Barabash:2020nck,PRITYCHENKO2025101694}.
The neutrinoless double beta decay (\ndbd) process,  $(A,Z) \to (A,Z+2) + 2e^-$, has never been observed and can only be induced by mechanisms beyond the Standard Model (BSM)~\cite{Agostini:2022zub}.

The detection of \ndbd\ would be a major breakthrough~\cite{PhysRevD.110.030001,Agostini:2022zub,Gomez-Cadenas:2023vca}, 
proving conclusively that the neutrino is a Majorana particle rather than a Dirac particle. This would distinguish the neutrino from other fermions, showing that it is its own antimatter partner. The discovery of \ndbd\ would also imply a new mechanism of mass generation, beyond the Higgs mechanism, to naturally explain the smallness of neutrino masses~\cite{Mohapatra:2007}. Moreover, the search for \ndbd\ is also a powerful, inclusive test of lepton number conservation. \ndbd\ produces two electrons -- particles of matter -- without the production of any antimatter, indicating lepton number violation (LNV). When combined with experiments that aim to precisely measure CP violation in the lepton sector, the discovery of LNV could help account for the matter-antimatter asymmetry of the universe~\cite{Fong:2012}. Indeed, from the perspective of BSM physics, LNV is as important as baryon number violation and should be pursued with the highest priority. 
We also note that \ndbd\ violates not only the total lepton number $L$ (which is an accidental symmetry of the Standard Model) but, more importantly, $B-L$ (where $B$ is the baryon number). $B-L$ is the only exact (non-anomalous) global symmetry of the Standard Model whose violation has not yet been observed. Therefore, experimentally investigating $B-L$ is of crucial importance.

The physically observable quantity in a \ndbd\  search is the decay rate $\Gamma^{0\nu}$. Usually, bounds on the half-life \thalf\ are quoted, where \thalf$\ =\ln(2)/\Gamma^{0\nu}$. The decay rate is set both by the BSM physics that enables \ndbd, the complex nuclear physics of the considered isotope and the phase space available for the decay. The dependence on the nuclear matrix elements (NMEs) and on the phase space make the expected \ndbd\ rate vary between different isotopes by multiple orders of magnitude. 
In order to compare the sensitivities of experiments studying different isotopes and to set well-defined experimental targets, we typically assume as baseline model the light-Majorana neutrino exchange~\cite{Agostini:2022zub}.
In this paradigm, the BSM physics driving \ndbd\ is neatly encapsulated in a single parameter called the effective Majorana neutrino mass, \mbb. Due to the complexity of the nuclear physics involved in the decay, the conversion between \thalf\ and \mbb\ is typically only known to within a factor of 2--3 ~\cite{Engel:2016xgb}. As a result, limits on and projected sensitivities of \mbb\ are typically reported as a range of values, reflecting the uncertainties in the nuclear physics calculations.

\subsection{Isotope choice}

The \ndbd\ signal is a peak in the summed-energy spectrum of the two emitted electrons, located at the Q-value (\Qbb) of the reaction. 
Given the long expected lifetimes -- of at least 10$^{25-26}$ yr -- the search for \ndbd\ relies on isotopes (which will be reviewed below) with relatively short predicted half-lives  and requires the use of as large an amount of isotope as possible.

In order to observe this rare process a detector must have high energy resolution, high efficiency and very low backgrounds~\cite{Agostini:2022zub,Giuliani:2012zu,Cremonesi:2013vla}.
High \Qbb\ values are a crucial advantage as the phase space factor $G_{0\nu}$
has a leading term proportional to \Qbb$^5$ and radioactive backgrounds tend to be lower at higher energies. In particular, \Qbb 's higher than 2.6~MeV are attractive, as the expected signal lies outside the bulk of the natural \G\ radioactivity. However, the isotope choice must cope with limitations imposed by the available detection technologies, isotope availability on the market, and overall project cost. As a result, some isotopes with \Qbb\,$> 2.6$~MeV, namely $^{48}$Ca, $^{96}$Zr, and $^{150}$Nd, are currently ruled out of viable and competitive large-scale experimental programs. Conversely, the isotopes $^{76}$Ge, $^{130}$Te, and $^{136}$Xe are widely studied thanks to the availability of well-established detector technologies and despite having \Qbb values in a region populated by natural \G\ emission. In fact, they currently provide the strongest limits on \mbb, i.e. \mbb\,$<$\,36--156\,meV from \Xe\ in KamLAND-Zen~\cite{KamLAND-Zen:2022tow}, \mbb\,$<$\,78--180\,meV from \Ge\ in GERDA~\cite{GERDA:2020xhi}, and \mbb\,$<$\,90--305\,meV from \Te\ in \cuore~\cite{CUORE:2021mvw}. Even at the completion of their experimental program, these projects will not cover the Inverted Ordering (IO) region for neutrino masses, corresponding to \mbb$\in[\sim20,\sim50]$\,meV\cite{Agostini:2022zub}.

In order to meet the challenge of next-generation \ndbd\ search experiments to probe the full IO region, a less well studied group of isotopes may be of interest. In particular, \Se, \Mo, and \Cd\ all have \Qbb\ values just above the bulk of environmental \G\ backgrounds. \Mo\ is a specially attractive candidate. It features \Qbb\,$=$\,3034\,keV and its highly favorable nuclear and phase-space factors yield an expected \ndbd\ rate up to an order of magnitude faster than the leading candidate isotopes (\Ge, \Xe, and \Te) for the same \mbb. The technology that enables the use of these alternative isotopes
is that of scintillating bolometers \cite{Poda:2021}, as shown by several small-scale demonstrators operated in the recent years: \Se\ in LUCIFER and \cupido~\cite{Azzolini:2018dyb,Azzolini:2019tta}, \Mo\ in LUMINEU~\cite{Armengaud:2015,Poda:2017d}, \cupidmo~\cite{Armengaud:2021,Augier:2022znx} and AMoRE~\cite{Alenkov:2019jis,AMoRE:2024cun}, and \Cd\ in an exploratory R\&D hosted by the CROSS and EDELWEISS facilities~\cite{CROSS:2023xdt,Helis:2020}.

\subsection{The scintillating bolometer technology}

Bolometers are ideal detectors to perform \ndbd\ searches~\cite{Fiorini:1983yj,Pirro2006,giuliani:2012a,Poda:2017c,Bellini:2018qhw}:
they feature an energy resolution at the few per-mil level,
a detection efficiency at the 70\%--90\% level,
and extremely low backgrounds due to the high radiopurity 
achievable in the crystals used as detectors~\cite{Arnaboldi2010,Armengaud:2017,Azzolini:2019nmi,S0217751X18430078}. 
Bolometers are somewhat unique in their versatility, allowing for the study of a wide range of \ndbd\ candidate isotopes, which includes \Ca, \Ge, \Se, \Mo, \Cd, $^{124}$Sn, and \Te.

A bolometer consists of a crystal absorber that converts an energy deposition from ionizing radiation into a temperature increase, coupled to a thermal sensor
that in turn converts the temperature into a voltage or current signal.
The typical mass of a crystal used in \ndbd\ searches is in the 0.1--1\,kg range~\cite{Fiorini:1983yj,Pirro2006,giuliani:2012a,Poda:2017c,Bellini:2018qhw}. Bolometers of this size must be operated at temperatures $\lesssim$20\,mK
so that the crystal heat capacity is low enough to provide high-amplitude signals with respect to the intrinsic noise sources. Historically, the limiting factor for scaling bolometric detectors has been the cryogenic infrastructure.
However, in the last decade the CUORE Collaboration has demonstrated the ability to operate stably $\sim$ 1\,ton bolometric detectors over several years~\cite{Alduino:2019,CUORE:2021mvw}.

As a result of a few decades of development, bolometric technology is now mature, and the last decade has witnessed extensive R\&D activities
to include a secondary readout of the scintillation light channel~\cite{Armengaud:2019loe, Barucci:2019ghi, Huang_2019, Casali_2019, Poda:2021, Alfonso:2022}. 
This has laid the groundwork for the implementation of a phased program to deploy increasingly sensitive detectors that take advantage of new and mature technologies, such as scintillating bolometers. This program will progress in tandem with anticipated advancements in millikelvin cryogenics, in synergy with growing demands from Quantum Information Science~\cite{Krinner}. These developments will enable further scaling of cryogenic infrastructures and consequently increased detector mass.

\begin{figure*}[htbp]
  \centering
  \includegraphics[width=0.7\textwidth]{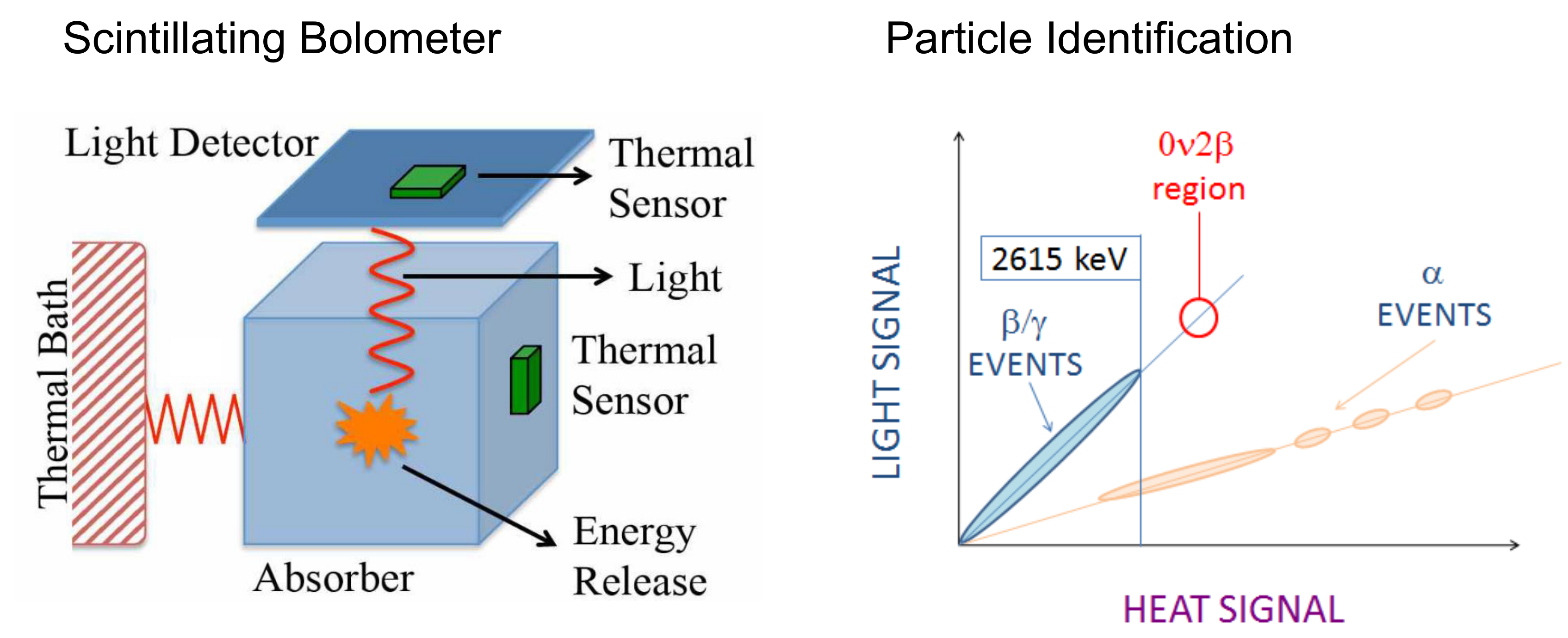}
  \caption{Left: A scintillating bolometer
    consists of two individual bolometric detector elements.
    The heat signal is read from the crystal absorber containing the \ndbd\ candidate isotope,
    while the light signal is detected on a second absorber sensitive to the scintillation light.
    The temperature rise in each absorber is measured by their respective thermal sensors.
    Right: Concept of \A\ particle rejection   by simultaneously measuring the light
    and heat signals and exploiting the different light yields of \A\ and \BG\ particles.}
  \label{fig:scintillating-bolometer-scheme}
\end{figure*}

A scintillating bolometer package (Fig. \ref{fig:scintillating-bolometer-scheme}) consists of two individual bolometric detector elements: a crystal absorber, with masses of the order of hundred grams, containing the \ndbd\ candidate isotope and having scintillating capabilities, and a thin wafer (usually in Ge or Si), with masses of the order of grams, facing the crystal absorber in order to collect its scintillation light. This device is a bolometric light detector (LD) capable of operating at mK temperatures where other devices -- such as SiPM or conventional photomultipliers -- fail or introduce insurmountable assembly and/or readout complications. We note that the scintillation light is proportional to the energy deposited; however, the conversion efficiency to scintillation depends on the type of particle interacting, with scintillation generally suppressed in heavy particles compared to electrons~\cite{TRETYAK201040}. The LD wafer is placed as close as possible to the scintillating crystal, but not in contact to avoid thermal cross-talk, and records the light signal. The scintillating crystal is a high-energy-resolution particle detector: the heat signal, following a particle interaction in its volume, provides an accurate measurement of the released energy in the form of heat. In combination with the heat signal, the LD is sensitive to particle species, and in particular is capable of distinguishing \A\ from \BG\ particles thanks to their different scintillation light yield (LY)~\cite{Pirro:2017,Poda:2021}. Therefore, the light signal can be used for Particle IDentification (PID), a powerful background suppression technique. As the LD is not in direct contact with the crystal, the light collection efficiency is sub-optimal. 
This calls for high-sensitivity LDs, especially if the chosen crystal is a modest scintillator. The relative LY is the experimentally observable quantity and depends on the scintillating crystal LY and on the light collection efficiency, which can be optimized by using anti-reflective coating on the LD\cite{Mancuso:2014paa}.

\subsection{The \cupid\ concept}

Building upon the successful cryogenic technology demonstrated by CUORE, and on several years of R\&D and optimization of scintillating bolometers embedding \ndbd\ candidates, \cupid\ aims at realizing a ton-scale experiment using \LMO\ scintillating crystals isotopically enriched in \Mo\ (Li$_2$$^{100}$MoO$_4$).
In fact, after preliminary investigations on \LMO\ ~\cite{Cardani:2013dia,BEKKER201638}, the technology developed in LUMINEU~\cite{Armengaud:2017,Grigorieva:2017} for this compound was applied to the CUPID-Mo demonstrator~\cite{Armengaud:2019loe,Armengaud:2021,Augier:2022znx}, 
proving \LMO\ as the optimal choice because of its acceptable intrinsic scintillator properties: this crystal provides a sufficient LY even in absence of doping ~\cite{Poda:2021} and the LY is compatible with the desired \A-background rejection~\cite{Armengaud:2019loe}. The demonstrated radiopurity of the crystals, which are grown from enriched material, satisfies the CUPID requirements~\cite{Augier:2023}.

\section{Detector description}\label{sec:detector}

\begin{figure*}[htbp]
  \centering
  \includegraphics[width=0.9\textwidth]{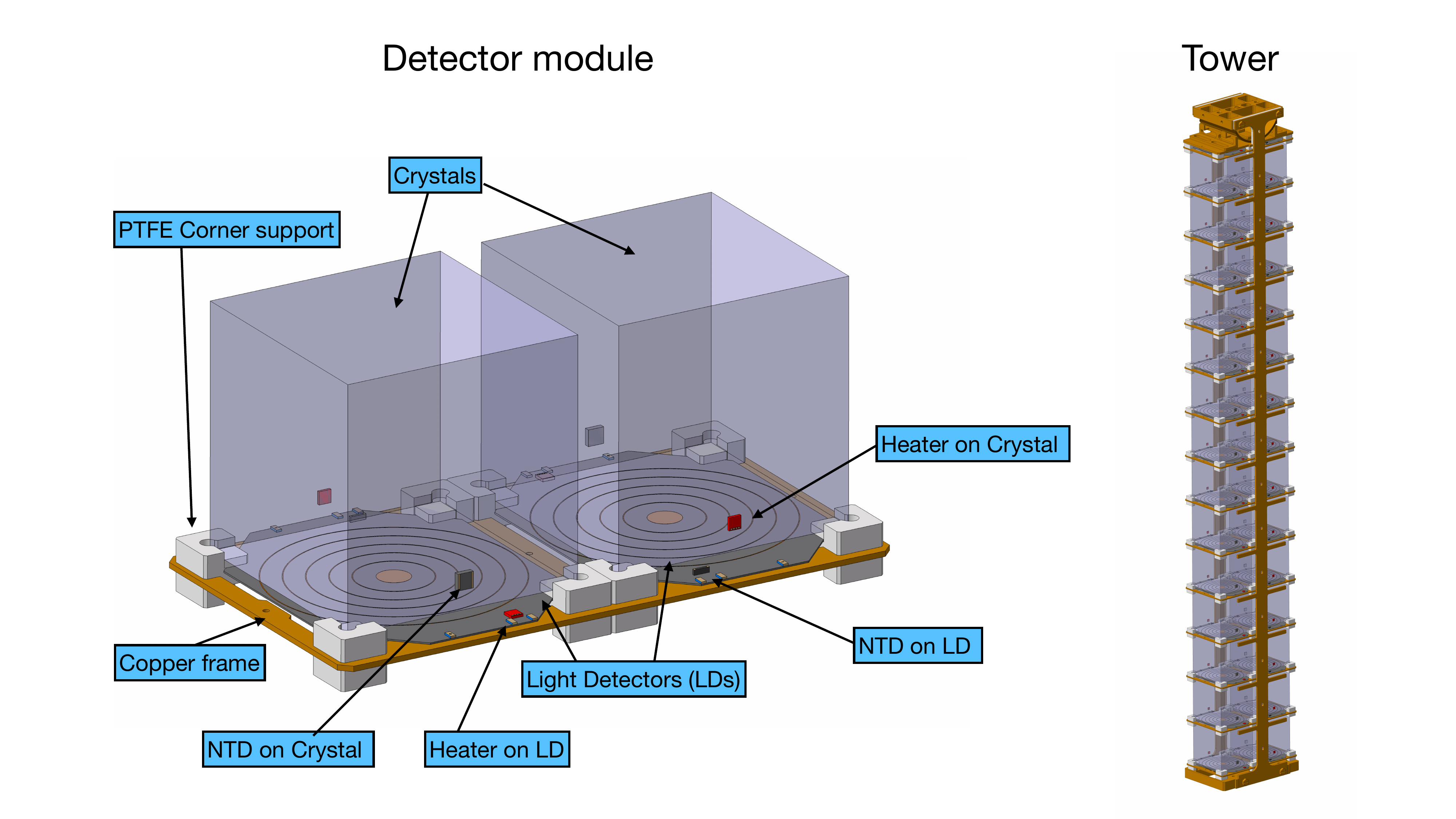}
  \caption{Left: schematic view of a single \cupid\ floor with two side-by-side detector modules,
    each consisting of a \LMO\ absorber (a cube with a side length of 45 mm) and a Ge light detector.
    Right: a single tower of 14 floors, or 28 detector modules.}
  \label{fig:CUPID-detector}
\end{figure*}

The \cupid\ detector design consists of a close-packed array of 1596 \enrlmo\ scintillating crystals
(Fig.~\ref{fig:CUPID-detector}) instrumented with the dual readout of heat 
and light to provide background tagging capabilities.
The crystals will be arranged in 57 towers,
each of them comprising 14 floors with two crystals per floor.

\cupid\ will be housed in the cryogenic facility presently hosting \cuore~\cite{CUORE:2021mvw,ADAMS2022103902}, at the Laboratori Nazionali del Gran Sasso (LNGS)
in Italy. It will benefit from the existing infrastructure, including the shielding against environmental radiation, the cryogenic system, the detector calibration systems, the vibration isolation system, and the muon-veto system. 

\subsection{Enriched \enrlmo\ bolometers}

The CUPID crystals will be grown from molybdenum enriched to $\geq$95\,\% in \Mo\ and cut into $45\times45\times45\,\mathrm{mm}^3$ cubes, corresponding to a mass of $\sim$280\,g per crystal, or a total \Mo\ mass of 240\,kg. The cubic geometry allows for a tightly packed detector design that optimally utilizes the experimental volume.
The crystal size was chosen as a compromise between two competing effects: larger crystals would maximize containment efficiency and minimize the number of readout channels required, whereas smaller crystals would decrease the rate of pile-up events induced by \nnbd\, which can contribute to background in the region of interest~\cite{Chernyak2012,Armatolb:2021}.

Each \enrlmo\ crystal will be instrumented with a Neutron Transmutation Doped (NTD) Ge thermistor~\cite{Haller1984}
for signal readout, and a Si heater for thermal gain stability control~\cite{Alfonso:2018a:pulser,Alessandrello:1998bf}. 
Enriched and natural \LMO\ crystals grown either with Czochralski and Bridgman techniques and having exactly the shape and size required by CUPID have already been tested in LNGS and in the Canfranc underground laboratory (LSC, Spain)~\cite{Alfonso_2023_JINST,Alfonso:2022,Armatol:2021}. Their bolometric performance is excellent, with energy resolutions and light yields approaching the CUPID requirements, which are respectively 5\,keV FWHM at 3\,MeV
and $\sim$0.35\,keV of energy collected by the LD for a 1\,MeV energy deposition from a \BG\ particle in the \LMO\ crystal. The capability to discriminate \A\ particles was proved.

\subsection{Light detectors} \label{sec:LD}

The LDs will be fabricated using octagonal-shaped,
300\,$\upmu$m-thick high-purity Ge wafers.
These wafers will be instrumented with NTD-Ge thermal sensors and heaters,
and  coated with a 70\,nm-thick SiO anti-reflective layer~\cite{Mancuso:2014paa} to enhance the photon absorption. Scintillation light from \LMO\ peaks at $\sim$600~nm at low temperatures~\cite{BEKKER201638}.
These types of photon detectors can reach signal-to-noise ratio (S/N) of $\sim$10 for
a scintillation signal induced by \BG\ deposition of 3\,MeV in the \LMO\ crystal, with the scintillating bolometer placed in an open structure without a reflective foil. This configuration is sufficient to provide a rejection of \A\ events better than 99.9\% with a 90\% \BG\ acceptance~\cite{Poda:2021},
and thus complying with the \cupid\ requirement in terms of \A\ background (Sec.~\ref{sec:bkgholder}).
These LDs also feature a rise-time (defined as the interval required for a pulse to increase from 10\% to 90\% of its maximum amplitude) as low as 0.5\,ms, which will be exploited to identify events induced by two \nnbd's randomly producing a pile-up event.

\begin{figure}
  \centering
  \includegraphics[width=\columnwidth,trim={0 0 9mm 0},clip]{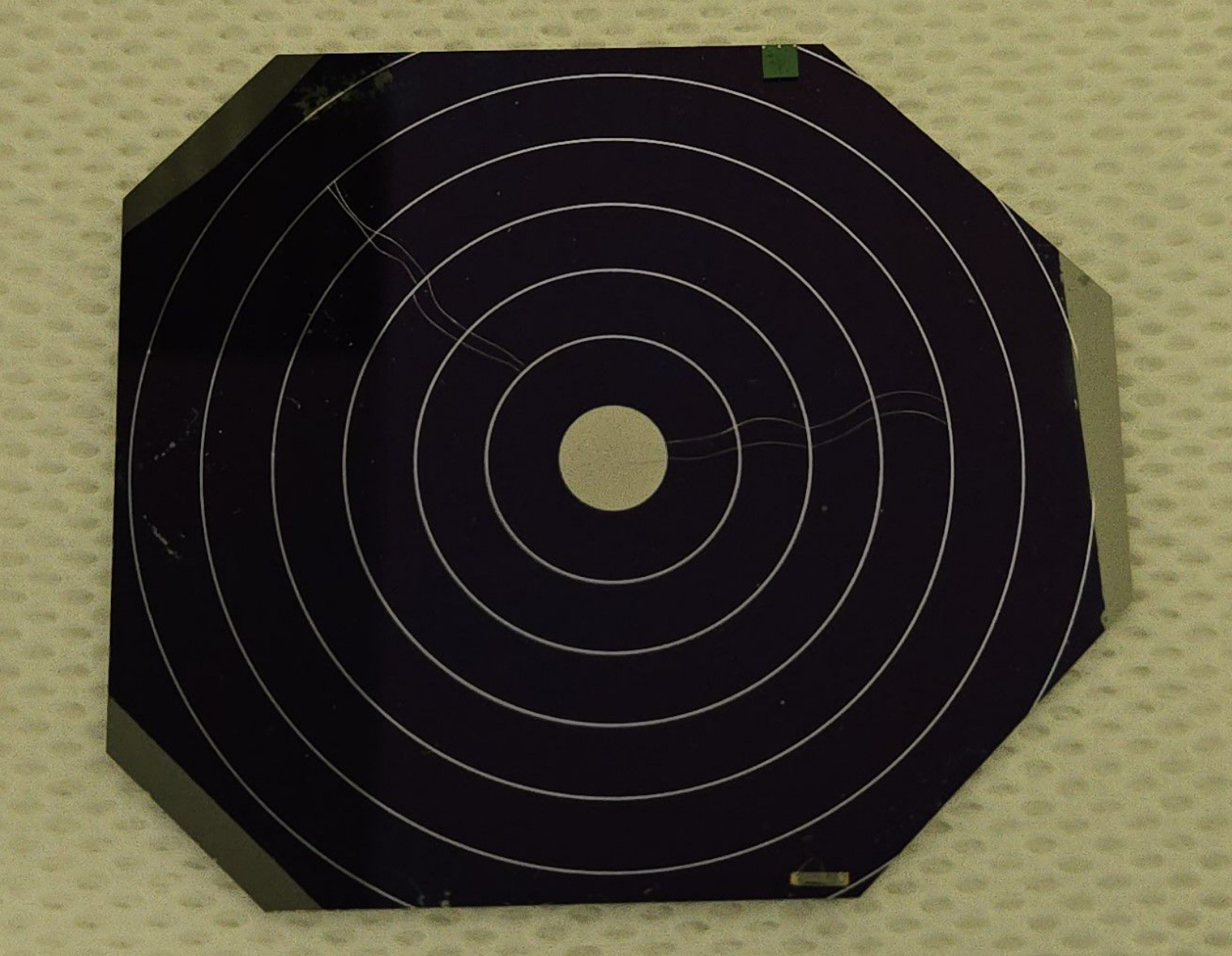}
  \caption { Example of an NTL Ge light detector without its copper housing, provided with two sets of concentric, annular interleaved Al electrodes. A SiO layer is deposited on both wafer sides to increase absorption of the scintillating photons emitted by the main \enrlmo\ crystal.}
  \label{fig:ntl-ld}
\end{figure}

The rejection of pile-up events will be further improved by instrumenting the LD with a set of Al electrodes, evaporated on the wafer surface. The Al electrodes are biased with a high voltage  $O$(100 V) producing a large electric field allowing for an enhanced S/N via Neganov-Trofimov-Luke (NTL) amplification~\cite{Neganov:1985khw,Luke:1988} (Fig.~\ref{fig:ntl-ld}).
This technology allows the LDs to achieve an effective S/N amplification of the order of 15~\cite{Novati:2019} and achieve a S/N of $\sim$150. Results obtained at LSC on several NTL LDs, developed for the CROSS \cite{Auguste_2024} and BINGO  \cite{ARMATOL2024169936} demonstrators,
proved an excellent performance both for \A~vs~\bega\ and pile-up rejection,
in compliance with the \cupid\ requirements.

\subsection{Detector assembly}

The detector structure is designed according to the following criteria:
the ease of assembly and cleaning of all structural parts,
the minimization of the time required for the detector assembly,
the minimization of dependence on mechanical tolerance,
and the reduction of all risks of failure in the assembly and wiring.
In the present design, both \enrlmo\ crystals and Ge wafers are instrumented with NTDs prior to the detector assembly. 
Each copper frame is equipped with eight PTFE corners, with two LDs positioned on top (Fig.~\ref{fig:CUPID-detector}). A small PTFE element covers the LDs to prevent direct contact with the \enrlmo\ crystals. Two \enrlmo\ crystals are then placed on top of the PTFE elements, and their weight compresses the PTFE, securely holding the LDs in place.

A full tower is composed of 15 copper frames with LDs in-between the 14 pairs of \enrlmo\ crystals.
With this design, all intermediate LDs will have line of sight to two crystals.
Finally, two copper spines are mounted vertically on the sides of each tower. Their aim is
to hold the tower weight and act as a support for the readout cabling,
composed of polyethylene naphthalate (PEN) bands instrumented with copper lines,
following the successful design of \cuore~\cite{Andreotti:2009zza}.
All connections between the NTDs, heaters or NTL amplification circuit and the copper strips
are performed with gold wire bonding.

In contrast to all previous bolometric prototypes~\cite{Armengaud:2017,Alduino:2016vjd,CUPID:2018kff,Armengaud:2019loe,Armatol:2021,Auguste_2024,ARMATOL2024169936,Alenkov:2019jis,Kim:2023},
the mechanical structure is entirely floating: each floor
is stacked on top of the previous one simply by gravity and not fixed to it with screws,
allowing us to significantly relax mechanical tolerances and to simplify the detector construction.
On the other hand, this innovative design requires dedicated tests to assess the bolometric performance in terms of thermal coupling and sensitivity to vibrations. The aim is to improve it, if necessary, with modifications that are as minimal as possible.
Therefore, in 2021 we started a campaign of tests on small \cite{Armatol:2021} and full-scale \cite{BDTP2025} prototypes.

Another fundamental difference with respect to the design of the previous demonstrators \cupido\ (in phase I) and \cupidmo\
is the absence of reflective foils around the scintillating crystals,
motivated by the need to reduce the possible background sources,
improve the capability to identify events that are not fully contained in one crystal, and simplify the overall detector design.
In order to maximize the light collection efficiencies,
we designed the detector holders so that the LDs are as close as possible
to the \enrlmo\ crystals~\cite{Armatol:2021,Alfonso:2022}.
This was achieved using rectangular LDs with cut-off corners
to allocate the space for the PTFE \enrlmo\-crystal holders and additional elements glued on the LD wafers for wire bonding~\cite{Alfonso_2023}.
We operated four LDs in a pulse-tube cryostat at the surface laboratory of IJCLab. In spite of a spring-based detector suspension used to mitigate vibrations~\cite{Mancuso:2014a}, the pulse-tube induced noise in this set-up was not mitigated as in the CUORE cryostat~\cite{Alduino:2019,Daddabbo:2018}.
Although the noise conditions were sub-optimal, all the LDs reached a baseline resolution between 70--90 eV~\cite{Alfonso_2023_JINST}, in compliance with the CUPID requirements of $<$ 100 eV baseline resolution.
Finally, we assembled a full-scale prototype (Fig.~\ref{fig:CUPID-detector}),
and operated it underground at LNGS. 
These R\&D runs proved that the mechanical and thermal properties satisfy the CUPID requirement for our \LMO\ bolometers,
allowing us to cool-down all detectors in $\sim$2 days. The FWHM energy resolution distribution across the 28 \LMO\ channels exhibited a median of 6.6 keV and a mean of 7.3 keV at 2615 keV. This outcome closely approaches the CUPID's target of 5 keV at 3034 keV, suggesting that the goal is indeed attainable with further refinement of noise reduction techniques. 
A detailed report on the outcome of these tests will be provided in a dedicated publication \cite{BDTP2025}. A further  measurement for the validation of the recently adopted NTL assisted LD technology within this or a slightly adapted structure is planned within spring 2025. 

\subsection{Data readout}
\label{sec:electronics}

\cupid\ will follow the read-out scheme of \cuore,
but will implement several improvements to achieve
a higher channel density, higher ADC resolution and sampling rates, and low noise for the LDs.
The front-end consists of room-temperature differential voltage preamplifiers~\cite{Arnaboldi:2009:pre,Arnaboldi:2017aek}
installed on top of the cryostat, inside a Faraday cage~\cite{Bucci:2017gew}.
We selected a new input JFET with higher transconductance and thus lower noise for a given power dissipation.
The preamplifiers for the LDs feature higher power consumption and lower noise (330\,mW and 1.5\,nV/$\sqrt{\text{Hz}}$ white noise)
than those for heat channels (180\,mW and about 3\,nV/$\sqrt{\text{Hz}}$).

The scheme of the NTD biasing circuitry will follow the one of \cuore, with an adjustable voltage applied to a pair
of custom large-value and high-aspect-ratio resistors (tens of G$\Omega$ range, with negligible low frequency noise coming from their small electric field per unit length~\cite{Arnaboldi:2002}) to generate the required bias current. For CUPID, we plan to increase the bias voltage up to 100~V, so that, at a given bias current, higher value resistors can be chosen, lowering their thermal parallel noise contribution. The thermal noise from load resistors was not impacting the energy resolution in CUORE~\cite{Adams:2022mod}, but it could be critical in CUPID especially for the LD readout, where we aim at the best possible S/N to control the random-coincidence background~\cite{Chernyak2012,Armatolb:2021}.

The power supply will include a low-noise commercial AC/DC stage
and a linear regulator~\cite{Carniti:2016a:linear} with higher Power Supply Rejection Ratio and low thermal drift and noise.

We also foresee an upgrade for the stabilization pulser boards~\cite{Alfonso:2018a:pulser}
by increasing the DAC resolution and expanding the firmware capabilities
to allow the injection of pulses over a DC signal for LDs or the injection of custom pulses for the study of pile-up rejection.

The NTL voltage supply will have a minimal impact on the overall wiring system, as LDs will be grouped and biased in parallel. While the exact grouping scheme is yet to be finalized, we will prioritize a pattern that prevents any given \enrlmo\ crystal from facing two LDs belonging to the same group. A basic configuration would involve two groups per tower, resulting in a total of 114 groups. This setup would require 114 individual wires, plus a few additional common wires for voltage reference. The NTL supply lines will be shielded from signal wires to minimize cross-talk. Each LD group will be connected through an RC filter to an external, commercial-grade power supply unit, characterized by low ripple output and capable of delivering up to 200\,V.

The back-end electronics, responsible for the filtering and digitization of the detector signals,
will adopt a custom solution~\cite{Carniti:2020:daq1, Carniti:2023:daq2}
that integrates a programmable anti-aliasing filter, and 24-bit ADCs,
with a sample rate of up to 25~ksps per channel.
A set of FPGAs produce a continuous data stream over 1\,Gbit/s Ethernet. 
The power consumption of this solution is a factor 5 lower than in \cuore, while occupying half the space. Compared to \cuore, this solution will allow us to enhance the precision of our time synchronization with real-world reference clocks.

The waveform sampling rate is planned to be set 1 kHz for the heat channels and 10 kHz for the LDs, which exhibit faster signal responses. This configuration will result in a data accumulation rate of 2.6 PB/yr, determining the storage requirements. Hardware digitizers will transmit data to several dedicated DAQ machines. These machines will record the data and synchronize it with the above-ground U-LITE (Unified LNGS IT Environment) system~\cite{Demin:2019}, which is an integrated infrastructure for scientific computing at LNGS. At U-LITE, a dedicated storage system will be implemented, where a near-real-time event reconstruction system will be operational.

\subsection{Calibration and light-detector regeneration}
\label{sec:calibration}

The heat channels will be calibrated following the \cuore~method, i.e., by deploying source strings immediately outside the cryostat, close to the detector region. Dedicated low-radioactivity polyethylene tubes will guide the strings, which will carry $^{232}$Th-loaded tungsten wires. The relatively long absorption length in \LMO\ (about 8~cm) for 2.6\,MeV photons, corresponding to the highest energy \G~line in the source, makes the calibration of even the most internal towers possible. Unlike in the \cuore~case, \Mo's \Qbb~is located at higher energies than all the $^{232}$Th-source lines. Therefore, we will occasionally perform calibrations with freshly produced $^{56}$Co sources~\cite{Augier:2023} --- which have a half-life of only 77.3\,days --- that provide \G~lines up to 3.4\,MeV~\cite{WANG1988791,DRYAK2008711}.

The energy calibration of the light channels can be very efficiently performed by exploiting the characteristic Mo X-rays~\cite{Armengaud:2019loe}, featuring two main lines groups at 17.4\,keV (K$_\alpha$) and 19.6\,keV (K$_\beta$). The X-rays are emitted by fluorescence when the \LMO\ crystals are irradiated during the calibration. The intensity of the source can be increased for LDs' calibration in order to improve the statistics in the Mo X-ray peaks.

A set of low-radioactivity optical fibers will be deployed in the detector volume, capable of providing close illumination to all the LDs. The fibers will be coupled to LEDs at room temperature which can emit light pulses at a wavelength close to the maximum of \LMO\ scintillation emission, i.e., $\sim$600~nm. Light pulses will be used to select the optimal bolometric operation points of the LDs, as well as to provide substantial continuous illumination during the so-called regeneration process, which consists of grounding the electrodes under strong LD wafer illumination~\cite{Novati:2019}. This operation will last a few minutes and will be performed once per month to remove the space charge accumulated in the Ge wafers that can lead to deterioration of the LD performance.

\section{Facility upgrade}

\cuore\ has pioneered the possibility of operating a large mass
of cryogenic detectors at 10--15\,mK temperature \cite{CUORE:2021mvw, ADAMS2022103902}.
The \cuore\ cryostat, at present, is the largest dilution refrigerator
ever built and operated worldwide,
in terms of cold mass and experimental volume.
The \cuore\ detector was cooled to base temperature in 2017
and never warmed-up to room temperature ever since, as of the end of 2023.
Despite its extremely successful operation, the \cuore\ cryostat must undergo
several upgrades to successfully host the CUPID detector.
These upgrades are driven by the CUPID design,
which entails a larger number of detectors and readout channels, leading to a more significant heat load for the cryogenic system. Additionally, they are informed by the data collected by \cuore\ regarding noise and its impact on the energy resolution of the detectors.
The upgrade is focused on the increase of cooling power on the 40\,K and 4\,K stages,
and  the reduction of vibrational noise necessary to improve the \LMO~energy resolutions and the LD S/N.

As described in Sec.~\ref{sec:detector}, the number of \cupid\ readout channels
is a factor of $\sim$3 larger than in \cuore.
The entire cabling, connecting the mixing chamber stage to the room-temperature electronics,
will be completely replaced.
As in \cuore, the cabling will consist of woven ribbon cables with twisted NbTi pairs~\cite{Andreotti:2009zza},
with strong thermal links to the cryostat stages
at 4\,K, 600\,mK, 50\,mK, and at the mixing chamber plate.
A possible increase of the thermalization at the 40\,K stage,
inside the pass-trough tubes that guide wiring from the room-temperature connection boxes
directly to the inside of the Inner Vacuum Chamber,
is under study to minimize the thermal load on the 4\,K plate.

To account for the increased thermal load, and consequently for the higher required cooling power,
we will install new pulse tubes (PTs). The PTs currently used in \cuore\
are PT415-RM from Cryomech~\cite{CRYOMECH415}, providing a cooling power of  1.35\,W at 4.2\,K.
The new Cryomech PT425-RM models ~\cite{CRYOMECH425}  provide a cooling power of 2.35\,W at 4.2\,K.
In order to account for possible failures, we will instrument the cryostat with a spare PT
to be turned on in case of need.
The PTs will be coupled through gas-gap heat switches, which allows to render the heat load from the spare PT (switched-off) negligible. In this configuration 3/4 PT425 are sufficient to operate CUPID, compared to the 4/5 PTs currently installed in \cuore.

Reducing vibrational noise is a crucial issue in any cryogenic detector experiment.
\cuore\ has shown that there are contributions to the energy resolution due to vibrations
induced both by the PTs and by environmental or human-induced sources.
The use of PT cryocoolers to replace the liquid helium bath in dilution refrigerators
has improved the duty cycle but brought an intrinsic vibration source in the experimental setup.
The pulsed pressure waves with the characteristic 1.4 Hz frequency
induce unavoidable vibrations in the experimental setups~\cite{Olivieri:2017}.
Many technologies have been developed in recent years to minimize the vibration
transfer or to dampen/cancel the vibrations induced by PTs.
The \cuore\ cryostat being a unique infrastructure in terms of size
and number of PTs requires dedicated developments to face this problem.
We will further refine a technique, already demonstrated by \cuore,
to suppress the PT-induced noise by actively tuning the PT phases~\cite{Daddabbo:2018}
to optimally cancel vibrations at the detector support plate.
In addition, new thermal connection between the PTs and the thermal stages
at 40\,K and 4.2\,K are a crucial upgrade. 
A possible improvement would consist in replacing the current Cu braids with 6N-purity Al connections, capable of minimizing the mechanical coupling while maintaining the same or higher thermal conductance with respect to CUORE.

Finally, a muon-veto system (MVS) based on plastic scintillators
will be installed to tag muon-induced events,
which are expected to contribute to the background.
The MVS will consist of a set of vertical scintillator panels
to be arranged around the CUPID cryostat and a set of horizontal panels below the cryostat \cite{MV2025}.
A single module for the MVS consists of a 2.5\,cm thick scintillating panel
(dimensions 1\,m by 0.5\,m) with embedded wavelength-shifting (WLS)
fiber and SiPMs as LDs.
The dark noise of SiPMs and environmental $\gamma$ rays,
the highest energy of which is 2615\,keV from the decay of
$^{208}$Tl, create background events for the muon-veto modules. These can be easily rejected by properly adjusting the trigger threshold, which we have tuned to keep false-positive rates such that the deadtime will be $<$1\%, while tagging $\sim $99\% of muons.
Additionally, the modules must be compact and fit within
the tight constraints of the infrastructure of the CUPID experiment.
The panels of the muon-veto modules will be made of plastic scintillator with embedded WLS fibers.

\section{Background}\label{sec:bkg}

The \ndbd\ signature for \Mo\ decay is an excess of events at \Qbb\,$=$\,3034\,keV.
We therefore evaluate the background index (BI) as the expected number of events,
induced by any background source, in a 30\,keV region of interest (ROI) around \Qbb,
normalized by the total detector mass, measurement livetime, and keV of energy.
This region is sparsely populated by events from natural radioactivity. The closest \G\ lines are from 3000 keV and 3054 keV photons emitted in $^{214}$Bi decays, both of which have a branching ratio at the 10$^{-4}$ level and are just outside the 30 keV ROI.
Our chosen ROI is nearly 6 times our goal energy resolution (FWHM\,=\,5\,keV).

The decay signature of a \ndbd\ event also has a distinguishing feature compared to most \G\ events. The energy deposition comes from electrons, which are fully contained within the crystal in $\sim$ 78\% of cases, whereas \G's are more likely to undergo Compton scattering and be absorbed in at least two different crystals.
This, in addition to the capability to perform PID are extremely powerful tools for background suppression.

To achieve a sensitivity that fully covers the IO mass range,
\cupid\ must reach a BI $\leq 10^{-4}$\,\ckky,
a level that is two orders of magnitude lower than that of \cuore\ in the same energy region.
In \cuore, most of the recorded events at 3\,MeV are produced by \A\ particles generated by decays of \U\ and \Th\ and their progeny occurring at the surface of passive materials building up the detector holder structure and facing the crystals.
This source, already identified in Cuoricino~\cite{CUORE:2003fpg} and \cuoreo\ \cite{CUORE0}, produces a nearly flat background that dominates the 3--4 MeV energy region. Hence, PID is the key strategy to reach the \cupid\ designed BI.

Table~\ref{tab:rate} shows the effect of PID in the demonstrator experiments \cupido~\cite{Azzolini:2022}
(26 ZnSe scintillating bolometers, 24 of which 95$\%$ enriched in $^{82}$Se)
and \cupidmo~\cite{Augier:2022znx} (20 \enrlmo\ scintillating bolometers, all of them 97$\%$ enriched in $^{100}$Mo).
In both cases, the counting rate in the ROI due to \A\ particles is reduced by one to two orders of magnitude until other backgrounds become dominant.
Though PID is not available in \cuore, its effect can be evaluated
on the basis of the background model~\cite{Adams:2024}.
We can notice from the last column of Table~\ref{tab:rate}
how the background would be much lower in \cuore\ than in the other experiments if PID were introduced with an \A\ tagging efficiency comparable to that achieved in the scintillating-bolometer demonstrators. 
This result is ascribable to the self-shielding and to the higher efficiency
of the anti-coincidence cut resulting from the larger number of detectors deployed in \cuore,
as well as to the higher radiopurity of its infrastructure.

\begin{table*}[htbp]
  \caption{Background counting rate recorded in past experiments in the \Mo\ ROI. It is to be noted that the close-material selection in terms of radiopurity was less stringent in CUPID-Mo with respect to CUORE and CUPID-0.
    Columns from 3 to 6 report:
    the rate of events before PID, releasing energy in just one detector (operation in anti-coincidence);
    the rate of \A\ events;
    the $\mu$ rate;
    the final rate after removing $\mu$ and \A\ events, and, for CUPID-Mo and CUPID-0,
    also the delayed coincidence events.
    The $\mu$ and \A\ rates are evaluated from MC-based background reconstructions, as is the BI after a hypothetical PID cut in CUORE. The muon rate in CUPID-Mo is estimated to be negligible due to the deeper location of this experiment in the Modane underground laboratory (France) combined with a muon-veto system hermetically surrounding the experiment.
  }
  \label{tab:rate}
  \centering
    \begin{tabular}{lccccc} 

    \toprule
    Experiment & Crystal     & BI before PID     & \A\ rate          & $\mu$ rate          & BI after PID       \\
               &             & [\ckky]           & [\ckky]           & [\ckky]             & [\ckky]            \\
    \midrule
    CUPID-Mo~\cite{Augier:2023} &\enrlmo      & 1.18 $\cdot 10^{-1}$  &   $1.15 \cdot 10^{-1}$                &                     & $2.7\cdot 10^{-3}$ \\
    CUPID-0~\cite{Azzolini:2019nmi} & Zn$^{82}$Se & $3\cdot 10^{-2}$  & $2.6\cdot10^{-2}$ & $1.5\cdot10^{-3}$ & $2.6\cdot 10^{-3}$ \\
    CUORE~\cite{Adams:2024} &TeO$_{2}$    & $1.4\cdot10^{-2}$ & $1.3\cdot10^{-2}$ & $4.1\cdot10^{-4}$   & $6.1\cdot10^{-4}$  \\
    \bottomrule
    \end{tabular}
\end{table*}

\begin{figure*}[htbp]
  \includegraphics[width=\textwidth]{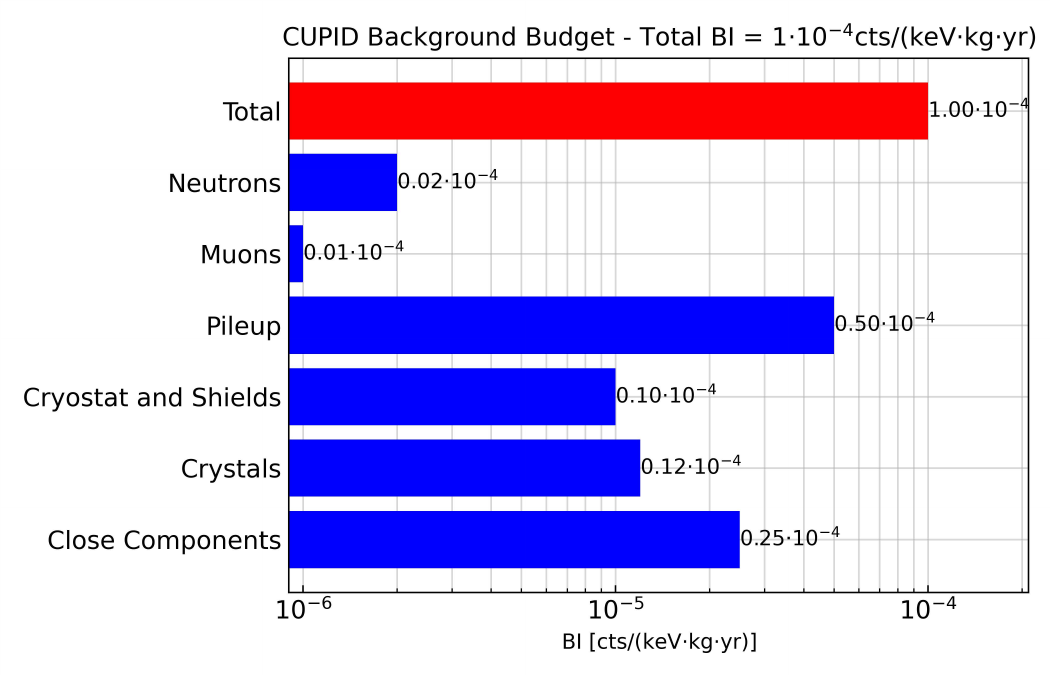}
  \caption{Breakdown of the predicted \cupid\ BI at the \Qbb\ of \Mo. 
    The predictions for the crystal contributions are derived from the \cupidmo\ background model~\cite{Augier:2023}.
    The pile-up estimates are based on the results from recent R\&D measurements~\cite{Ahmine2023}.
    The detector components predictions are extrapolated from the \cuoreo\ background model~\cite{CUORE0} with the addition of the light detectors,
    those of the \cuore\ infrastructure are obtained from \cuore~\cite{Adams:2024},
    and those from external radiation from dedicated MC simulations.}
    \label{fig:CUPID-BB}
\end{figure*}

In the evaluation of the \cupid\ background budget,
we have grouped sources that may produce background events in the ROI
into a set of uncorrelated categories and, for each category,
have fixed an upper level that cannot be exceeded.
These upper levels sum at the \cupid\ BI goal as shown in Fig.~\ref{fig:CUPID-BB}.
Below, we briefly describe the sources belonging to each of the categories,
the present knowledge we have on their intensity and the steps that will lead
to their reduction to or below the allowed threshold, and finally the motivation
for choosing that specific threshold. The results and projections rely
on the data-driven background models developed for \cuore~\cite{Adams:2024}, \cuoreo~\cite{CUORE0}
and \cupidmo~\cite{Augier:2023} experiments
and on a GEANT4 simulation of \cupid . This simulation is based on the one performed for \cuore\ with the required modifications to account for the new detector material and new overall design. 

\subsection{Environmental radiation (\G, n, and $\mu$)}

Photon, neutron and muon fluxes, though dramatically reduced
by the LNGS mountain overburden, can induce a sizable background rate in the ROI.
The \cuore\ cryostat has a massive external passive shield consisting of 18\,cm
of polyethylene and 2\,cm of H$_3$BO$_3$ powder to moderate and absorb environmental neutrons, respectively,
and at least 25\,cm of lead in all directions to suppress environmental \G\ radiation.
This shielding, together with the anti-coincidence cut, ensures that environmental radiation
is a subdominant contribution for \cuore.
In \cupid\ however, where the target background is much lower,
a further suppression of the neutron- and $\mu$-induced events is necessary.
We set as a target contribution for external radiation 10\% of the goal BI, i.e. $10^{-5}$\,\ckky.
We chose this value so that it is comparable to the background coming from the infrastructure (see Sec.~\ref{sec:infrastructurebkg}),
and subdominant with respect to the total BI.

Simulations provide estimates for the expected $\mu$ rates:
1.8\,muons/hour in the CUPID crystals, and 7.0\,muons/hour in the lead shield.
They also show that the proposed muon-veto system would have a muon rejection efficiency at trigger level
of $\sim$98.0\%,
achieving a background index in the ROI of $4.6\cdot10^{-6}$\,\ckky.

For neutrons, results of Monte Carlo (MC) simulations
indicate the need to expand the neutron shield with an additional 10\,cm layer of polyethylene
to fully moderate neutrons of up to 10\,MeV of energy
and prevent the occurrence of (n,\G) reactions in Cu and Mo isotopes.
In our first schematic studies of such a shielding hermetically surrounding the cryostat the background in the ROI is suppressed to $0.02\cdot10^{-4}$\,\ckky.

\subsection{Cryostat and shields}\label{sec:infrastructurebkg}

In \cupid, the cryogenic infrastructure will remain mostly unchanged with respect to \cuore.
Such infrastructure includes the dilution refrigerator with its system of nested thermal shields
and the radiation shields. The innermost cryostat thermal shield contributes to the background
via the emission of \A\ and \B\ particles produced by radioactive contaminants
on its inner surface, which has a direct line of sight to the detectors.  In \cuore, the innermost shield consists of a copper cylinder,
internally paved with copper tiles.
Despite the very efficient rejection of \A\ induced events achieved with PID,
surface contaminants remain a dangerous source of background, as they can contribute, at a much lower level but still challenging for \cupid , through \BG\ emissions. Their suppression strongly depends on the use of special surface treatments as those already developed for \cuore~\cite{ALESSANDRIA201313}.

Since the \enrlmo\ array will be enclosed in a copper shield with few mm thickness,
an additional contribution, albeit smaller,  to the background from the cryogenic infrastructure is represented by \G\ rays produced by
\Bi, which emits a few low-intensity \G\ rays with energy larger than \Qbb,
and \Tl, which emits two \G\ rays in coincidences with 2615 and 583\,keV,
which could lose a small fraction of energy via Compton scattering in the passive materials
producing an event of $\sim$3\,MeV in a single crystal.

We fix our goal for the infrastructure contribution to 0.1$\cdot 10^{-4}$\,\ckky,
accounting for the possibility to further reduce it by replacing the innermost thermal shield of the cryostat
and the internal lead shield with new, ultra-clean copper and by replicating the same cleaning to the shield's copper surfaces.

\subsection{Detector holder - close components }\label{sec:bkgholder}

The detector holder is the mechanical structure, made of copper and PTFE, that supports
the \enrlmo\ crystals, as well as the cabling, made of copper and PEN.
These components provide a twofold background contribution:
their bulk contaminants -- homogeneous in the material volume --
can contribute via \G\ radiation, while their surface contaminants
-- introduced during machining or exposure to radon and concentrated on the surface --
can contribute via \A\ and \B\ particles.
From the \cuore\ experience we expect that a proper material selection
can guarantee a sub-dominant background contribution from bulk contamination,
while surface contamination is more problematic.
We set $0.25\cdot10^{-4}$\,\ckky\ as a target threshold for the detector holder background contribution,
which is a factor 2 lower than the value measured in \cuore.
To achieve this goal, the surface contamination activity
must be at the level of few nBq/cm$^2$,
a value that is well-beyond the sensitivity of current measurement techniques.
In \cuore, we have developed a cleaning procedure~\cite{ALESSANDRIA201313}
that removes material at the surface with ultra-clean reagents.
This technique is extremely efficient on flat surfaces,
but less effective on components with non-trivial shapes.
In \cupid, we have radically changed the design of the copper frames
in order to reduce the machining operations,
and succeeded in designing copper frames that can be produced
exclusively by lamination, bending and laser cutting,
thus fully avoiding the need to mill or drill.
In addition, we are actively working on a setup dedicated to the measurement
of surface contaminants~\cite{Benato:2023ijp}.

\subsection{\enrlmo\ crystals: bulk and surface}

Bulk and surface contamination of the \enrlmo\ crystals
in \U\ and \Th\ are inferred from the \cupidmo\ background model~\cite{Augier:2023},
and contribute to the BI with $0.12\cdot10^{-4}$\,\ckky. 
This value is dominated by surface contaminants and is a factor $\sim$4 lower than in \cupidmo\
thanks to the higher granularity of \cupid, and to the absence of reflective foils,
yielding a higher anti-coincidence efficiency.

Targeting zero exposure to air and following the measures similar to those adopted for \cuore\ and not strictly followed in \cupidmo , the crystals will be stored underground in N$_2$ flushed storage containers and brought above ground for their final instrumentation in dedicated partially automated assembly stations with controlled N$_2$ atmosphere. 

\subsection{\enrlmo\ crystals: \nnbd\ pile-up events}

The relatively slow response time of bolometers can lead to a background
from accidental pile-up of events in a single crystal,
when two events that occur close enough in time are not resolved,
but reconstructed as a single event at their summed energy.
Pile-up events can be produced by any active-enough source, and have a rate that is the product of the contributing sources' rates.
In \cupid, the source that is expected to dominate the detector counting rate
is the \nnbd\  of \Mo . This decay has a half-life of $7.1\cdot10^{18}$~yr~\cite{Augier:2023b} and results in an event rate of ~2.6 mHz within a single \enrlmo\ crystal with the size chosen for \cupid. The impact of this source can be appreciated considering that this rate would produce a BI of the order of $3 \cdot10^{-4}$~\ckky\ with a time resolution of 1~ms~\cite{Chernyak2012,Ahmine2023}. 

The rejection of pile-up events becomes increasingly difficult as events gets closer in time.
To a first approximation, the parameters that determine the ability
to identify pile-up events are the detector rise-time and the S/N.
While the time response of \lmo\ crystals is too slow to efficiently reject pile-up events~\cite{Armatolb:2021},
our strategy is to suppress them using LDs~\cite{Chernyak2016,Ahmine2023}.
To reach the $10^{-4}$\,\ckky\ background goal,
the contribution from pile-up events should be $\leq0.5\cdot10^{-4}$\,\ckky.
\cupid\ can reach this goal by implementing the NTL amplification
in the LD, as discussed in Sec.~\ref{sec:LD}.

\section {Sensitivity}
\label{sec:sensitivity}

Considering the experimental parameters explained so far and summarized in Table~\ref{tab:pars},
we use a frequentist approach to evaluate the \cupid\ sensitivity
both in terms of \ndbd\ half-life  and \mbb \cite{Sensitivity2025}.
In particular, we compute the discovery sensitivity,
defined as the value of \thalf\ for which \cupid\ has a 50\% probability
to obtain a signal excess with a significance of $\geq3\sigma$,
and the exclusion sensitivity, defined as the median of $90\%$ exclusion limits
obtained under the assumption of no signal.
In both cases, we convert the sensitivity in \thalf\ into a sensitivity in \mbb,
using a set of NMEs computed with different
nuclear models~\cite{LopezVaquero:2013yji,Hyvarinen:2015bda,Song:2017ktj,Simkovic:2018hiq,Deppisch:2020ztt,Barea:2015kwa}.

We model our energy spectrum with a flat background and a Gaussian signal
in a 140-keV region around \Qbb: this model is used both for generating toy-MC experiments,
as well as a fitting function.
We define our test statistics as
\begin{equation}
  t(\Gamma) = -2 \ln{\frac{H_0}{H_1}} = -2 \ln{\frac{\mathcal{L}(\Gamma,\hat{\nu})}{\mathcal{L}(\hat{\Gamma},\hat{\nu})}},
\end{equation}
where $H_0$ is the null, background-only hypothesis,
and $H_1$ is the alternative, signal-plus-background hypothesis,
while $\mathcal{L}$ is the likelihood function that is profiled
on the parameter of interest namely the \ndbd\ rate $\Gamma=ln (2)/T_{1/2}^{0\nu}$,
while fixed to the best-fit values for all nuisance parameters $\nu$.
To account for the fact that the decay rate can only take non-negative values,
we adjust the test statistics by evaluating the likelihood at zero
for all situations where $\hat{\Gamma}$ is negative.

\begin{table}[htbp]
  \caption{Experimental parameters used for the evaluation of the exclusion and discovery sensitivity.}
  \label{tab:pars}
  \centering
  \begin{tabular}{rl}
    \toprule
    Parameter & Value \\
    \midrule
    Total mass               & 450\,kg \\
    Enrichment fraction      & 95\%    \\
    Isotope mass             & 240\,kg \\
    Containment efficiency   & 78\%    \\
    Selection efficiency      & 90\%    \\
    Energy resolution (FWHM) & 5\,keV  \\
    Background Index         & $10^{-4}$\,\ckky \\
    Livetime                 & 10\,yr  \\
    \bottomrule
  \end{tabular}
\end{table}

\subsection{Discovery sensitivity}

Searching for a discovery is equivalent to performing a hypothesis test with:
\begin{equation}
  t_P(0) = -2\ln{\frac{\mathcal{L}(0)}{\mathcal{L}(\hat{\Gamma})}},
\end{equation}
where a claim of discovery (or evidence) can be made if the test statistic
is greater than some cut-off.
To ensure the correct coverage, we numerically produce the test statistic distribution
using toy-MC experiments for different values of the injected signal strength,
and compute the background-only p-value ($p_b$):
\begin{equation}
  p_b = \int_{t_P}^{\infty} f\left( t_P(0) | \Gamma=0 \right) dt,
\end{equation}
where $t_P$ is the value of the test statistic observed in each toy-MC experiment,
and $f(t_P|\Gamma)$ is the distribution of test-statistic for toy-MC experiments
generated with the background-only model.
A discovery can be claimed if $p_b$ is smaller than some cut-off,
that we set to 0.14\%, corresponding to a $3\sigma$ evidence.

With a 10\,yr livetime, which is realistically extrapolated from the operational experience gained with CUORE and the planned refurbishment of the cryostat, \cupid\ will reach a $3\sigma$ discovery sensitivity
of $\hat{T}_{1/2}^{0\nu} = 1\cdot10^{27}$\,yr.
This value corresponds to a set of different \mbb\ values.
The lower value is obtained with the EDF model~\cite{LopezVaquero:2013yji}, and corresponds to 12\,meV.
The upper value of this range can be obtained by the QRPA and IBM models~\cite{Barea:2015kwa,Hyvarinen:2015bda}
and corresponds to 21\,meV.

\begin{figure}[htbp]
  \centering
  \includegraphics[width=\columnwidth]{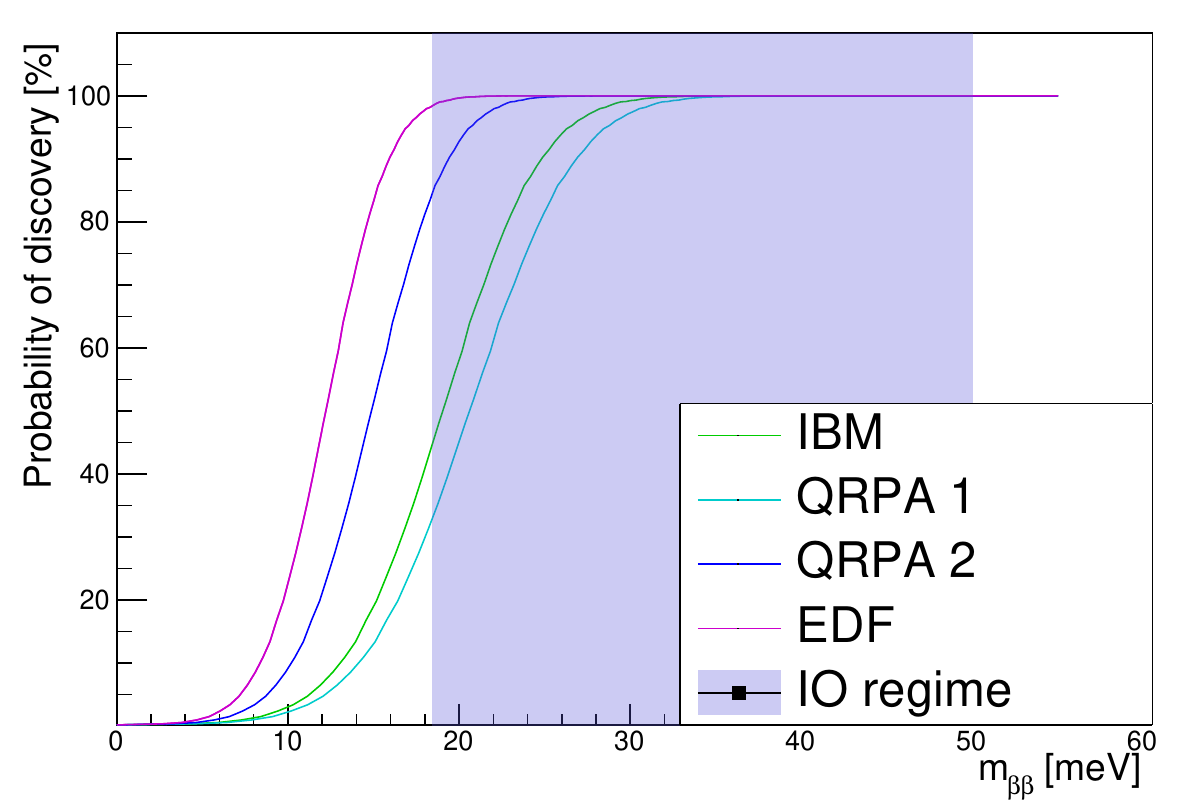}
  \caption{Discovery probability as a function of \mbb, assuming different values
    of NME~\cite{LopezVaquero:2013yji,Barea:2015kwa,Hyvarinen:2015bda,Song:2017ktj,Simkovic:2018hiq,Deppisch:2020ztt}.
    The shaded area corresponds to the allowed \mbb\ range in the inverted ordering,
  assuming $m_{\text{lightest}}\lesssim10$\,meV.}
  \label{fig:disc_mbb}
\end{figure}

Finally, we use the output of the discovery sensitivity to extract the discovery probability
as a function of \mbb. The results are reported in Fig.~\ref{fig:disc_mbb}.

\subsection{Exclusion sensitivity}

\begin{figure}[htbp]
  \centering
   \includegraphics[width=1.\columnwidth]{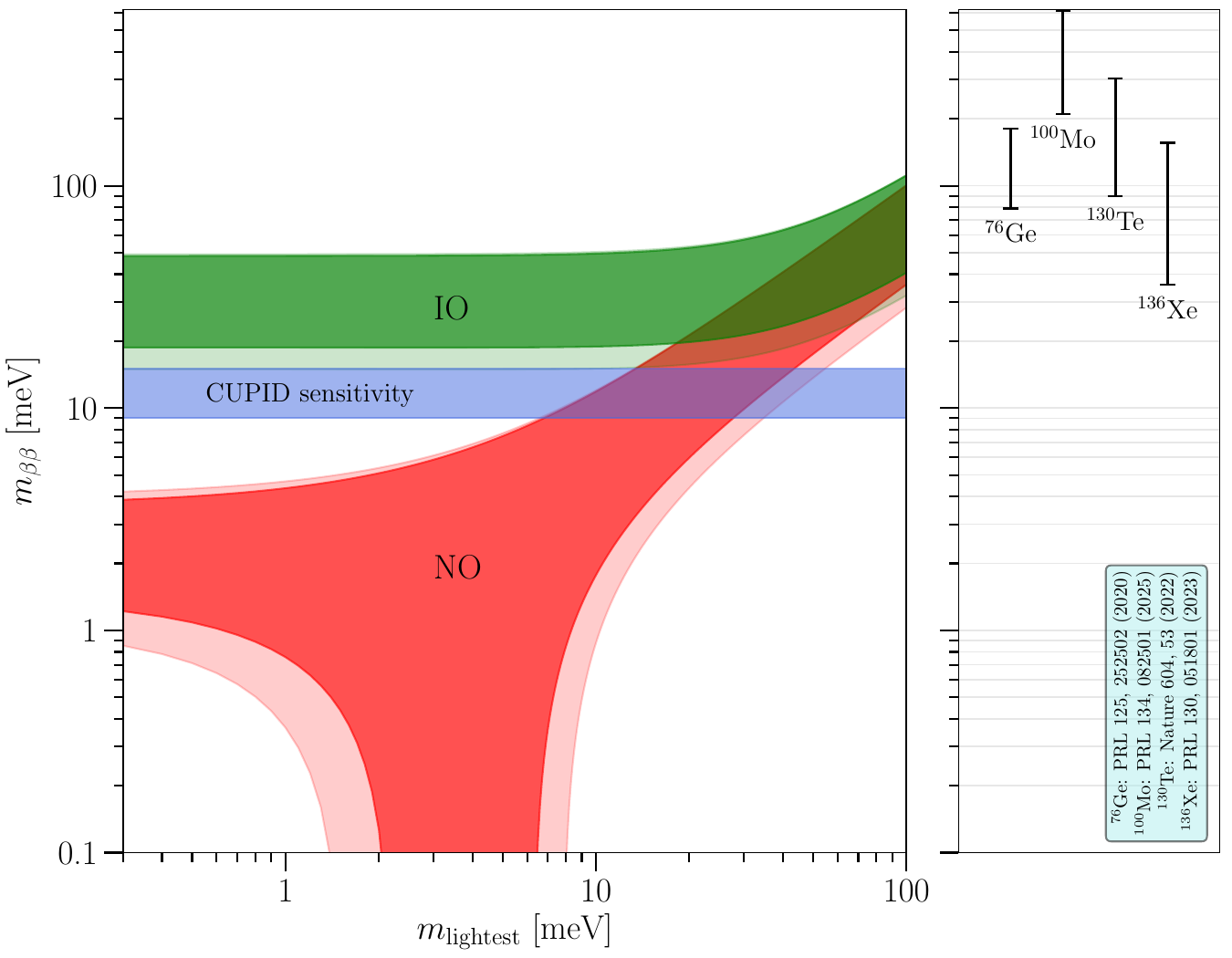}
  \caption{Constraints on the effective Majorana neutrino mass ($m_{\beta\beta}$)
    as a function of the lightest neutrino mass ($m_\mathrm{lightest}$).
    The dark regions represent the predictions on $m_{\beta\beta}$
    based on best-fit values of neutrino oscillation parameters for the Normal and Inverted
    Ordering cases (NO and IO); the shaded regions indicate the $3\sigma$-ranges
    calculated from the oscillation-parameter uncertainties~\cite{Capozzi:2021fjo}.
    The horizontal band indicates the projected CUPID sensitivity,
    corresponding to a half-life of \ce{^{100}Mo} greater than $1.8\cdot10^{27}$ yr,
    assuming a set of NME~\cite{LopezVaquero:2013yji,Barea:2015kwa,Hyvarinen:2015bda,Song:2017ktj,Simkovic:2018hiq,Deppisch:2020ztt}.
    On the side panel, the most recent limits from the isotopes
    \ce{^{76}Ge}~\cite{GERDA:2020xhi}, \ce{^{100}Mo}~\cite{amore2024new},
    \ce{^{130}Te}~\cite{CUORE:2021mvw} and \ce{^{136}Xe}~\cite{KamLAND-Zen:2022tow} are reported. 
		(Updated results for these isotopes, except $^{100}$Mo, are reported in preprints \cite{Adams:2024search,Abe:2024search,Acharya:2025first}.)
  }
  \label{fig:mbb_ml}
\end{figure}

To extract the frequentist exclusion sensitivity, we use the same approach as for the discovery,
but invert the hypothesis test to compare a null hypothesis $\Gamma=S$
to an alternative $\Gamma\neq S$.
We compute the distribution of $f(t_P(\Gamma)|\Gamma)$ using toy-MC experiments to obtain the p-value
\begin{equation}
  p_\mu(\Gamma) = \int_{t_P}^{\infty} f\left( f_P(\Gamma) | \Gamma \right)dt,
\end{equation}
then compute the 90\% upper limit for each toy-MC as the interval of $\Gamma$ values
with $p_\mu(\Gamma)>0.1$.
As a result, we obtain an exclusion sensitivity of \thalf\,$>1.8\cdot10^{27}$\,yr at 90\% confidence level.
In terms of the effective Majorana neutrino mass, this becomes \mbb\,$<$\,9--15\,meV.
The range of \mbb\ values that can be excluded by \cupid\ is shown in Fig.~\ref{fig:mbb_ml}.

CUPID considers deploying the detector in two phases, enabling early data collection. The evolution of sensitivity over time in different scenarios is discussed in Ref. \cite{Sensitivity2025}.

\section{Conclusions and outlook}

\cupid\ is a next-generation experiment to search for \ndbd\ 
and to perform other rare-event studies ---\nnbd\ spectral shape, Lorentz and CPT violation, \ndbd\ with Majoron(s), bosonic neutrinos and other phenomena \cite{Auger:2024} --- using scintillating bolometers enriched in \Mo. 
Its goal is to search for \ndbd\  with a discovery sensitivity
covering the full neutrino mass regime in the IO scenario.  CUPID will simultaneously probe a large portion of the Normal Ordering (NO) regime~\cite{Agostini:2017jim} with lightest neutrino mass larger than 10\,meV. 
This effective Majorana neutrino mass sensitivity corresponds to a \Mo\ \ndbd\ half-life of \thalf\,$=10^{27}$\,yr.

To achieve its science goals, \cupid\ uses two approaches.
First, while \cuore\ has demonstrated the infrastructure requirements with the candidate isotope \Te,
switching from \Te\ to \Mo\ moves the ROI above 2615\,keV,
significantly dropping the \G\ background by over an order of magnitude with respect to \cuore.
Second, the use of \LMO\ scintillating bolometer technology
allows the near complete rejection of the \A\ background
via the different light yields for \A\ and \BG\ interactions of the same energy.

\cupid\ construction activities have already started, including preparations for the cryostat upgrade and the setup of the detector assembly infrastructure. Based on current planning, data taking is expected to begin in the very early years of the next decade

The scintillating bolometer technology based on \LMO\ crystals highly enriched
in \Mo\ is scalable to ton-scale isotopic masses beyond the \cupid\ baseline design described here.
The modular nature of the \cupid\ crystal array allows us to conceive a phased future program.
An expansion to 1~metric ton of \Mo\ in a larger cryostat or in multiple experimental setups,
and a more aggressive background target, would improve the half-life sensitivity well beyond $10^{27}$\,yr 
or a \mbb\ sensitivity well below 10\,meV, 
probing deep into the NO regime. 

One of the unique advantages of the scintillating bolometric technology utilized by CUPID
is its possible application to several favorable candidates.
Competitive experiments can be performed using crystals of \enrZS,
\enrlmo, and \enrcwo, or a combination of them~\cite{giuliani:2018a}, if they can all be grown with the required high radio-purity.
These multiple searches can be performed by simply changing the crystals
while keeping the same detector configuration, assembly procedures,
cryogenic infrastructure, readout electronics, data-acquisition system, and analysis tools.
This becomes a crucial capability in the event of the discovery of a \ndbd\ signal,
at which point the measurement in the same detector setup and background environment
but with different isotopes  would provide the means for an unambiguous test of the signal and the means to begin to probe the underlying physics of \ndbd.

\subsection{Acknowledgments}

The CUPID Collaboration thanks the directors and
staff of the Laboratori Nazionali del Gran Sasso and
the technical staff of our laboratories. This work
was supported by the Istituto Nazionale di Fisica
Nucleare (INFN); by the European Research Council (ERC) under the European Union Horizon 2020
program (H2020/2014-2020) with the ERC Advanced
Grant no. 742345 (ERC-2016-ADG, project CROSS)
and the Marie Sklodowska-Curie Grant Agreement No.
754496; by the Italian Ministry of University and Research (MIUR) through the grant Progetti di ricerca di Rilevante Interesse Nazionale (PRIN)  grant no. 2017FJZMCJ and grant no. 2020H5L338; by the US National Science Foundation under Grant Nos. NSF-PHY-1401832, NSF-PHY-1614611, NSF-PHY-2412377 and NSF-PHY-1913374; by the French Agence Nationale de la Recherche (ANR) through the  ANR-21-CE31-0014- CUPID-1; by the National Research Foundation of Ukraine (Grant No. 2023.03/0213). This material is also based upon work supported by the US Department of Energy (DOE) Office of Science under Contract Nos. DE-AC02-05CH11231 and DE-AC02-06CH11357; and by the DOE Office of Science, Office of Nuclear Physics under Contract Nos. DE-FG02-08ER41551, DE-SC0011091, DE-SC0012654, DE-SC0019316, DE-SC0019368, and DE-SC0020423. This work was also supported by the Russian Science Foundation under grant No. 18-12-00003. This
research used resources of the National Energy Research Scientific Computing Center (NERSC). This
work makes use of both the DIANA data analysis and
APOLLO data acquisition software packages, which
were developed by the CUORICINO, CUORE, LUCIFER and CUPID-0 Collaborations.

\bibliographystyle{spphys}       



\end{document}